\documentclass[prd,nobibnotes,showpacs,onecolumn,nofootinbib]{revtex4}
\usepackage{graphics,color,array,dcolumn}
\usepackage{calc}

\usepackage{amsmath}
\usepackage{amssymb}
\usepackage{xspace}
\newlength{\figurewidth}
\usepackage{graphicx}
\definecolor{joerg}{rgb}{1.0,0.0,0.0}

\newcommand{\eq}{\begin{equation}}
\newcommand{\feq}{\end{equation}}

\allowdisplaybreaks[1]

\newcommand{\be}{\begin{equation}}
\newcommand{\ee}{\end{equation}}

\newcommand{\R}{{\mathcal{R}}}

\makeatother

\begin{document}
\setlength{\figurewidth}{\columnwidth}

\title{Renormalization Group Flow in Scalar-Tensor Theories. II}
\author{Gaurav Narain}
\email{narain@sissa.it}
\affiliation{SISSA, Via Beirut 2-4, Trieste 34151, Italy, and INFN, Sezione di Trieste, Italy}
\author{Christoph Rahmede}
\email{c.rahmede@sussex.ac.uk}
\affiliation{Department of Physics and Astronomy, University of Sussex, Brighton, East Sussex, BN1 9QH, UK}
\pacs{04.60.-m, 11.10.Hi}

\begin{abstract}
We study the UV behaviour of actions including integer powers of scalar curvature
and even powers of scalar fields with Functional Renormalization Group techniques.
We find UV fixed points where the gravitational couplings have non-trivial values
while the matter ones are Gaussian. We prove several properties of the linearized
flow at such a fixed point in arbitrary dimensions in the one-loop approximation
and find recursive relations among the critical exponents. We illustrate these results
in explicit calculations in $d=4$ for actions including up to four powers
of scalar curvature and two powers of the scalar field.
In this setting we notice that the same recursive properties among the critical exponents,
which were proven at one-loop order, still hold, in such a way that
the UV critical surface is found to be five dimensional. We then
search for the same type of fixed point in a scalar theory with minimal coupling
to gravity in $d=4$ including up to eight powers of scalar curvature.
Assuming that the recursive properties of the critical exponents still hold,
one would conclude that the UV critical surface of these theories is five dimensional.
\end{abstract}

 \maketitle

\section{Introduction}

In \cite{Gaurav} scalar-tensor theories were studied where the purely gravitational part
was given by the Einstein-Hilbert action. Here we generalize those results by including
higher curvature terms. The main aim of our analysis is to understand if gravity remains
asymptotically safe \cite{Weinberg,Weinberg:2009ca} under the inclusion of some matter
component. Results about the renormalizability of gravity can depend crucially on the
inclusion of matter. Already in the first one-loop calculations \cite{Veltman,Deser}
it was shown that pure gravity is one-loop renormalizable
but becomes one-loop nonrenormalizable in the presence of matter.
In the context of the search for asymptotic safety, it was shown in \cite{Perini1}
that the position and even the existence of a nontrivial gravitational fixed point
in the Einstein-Hilbert truncation is affected
by the presence of minimally coupled matter fields.
In \cite{cpr2} we showed that this is also true when higher derivative gravitational
terms are present.
In \cite{GrigPerc,Perini2,Gaurav} the effect of gravity on scalar interactions
was studied, assuming the Einstein-Hilbert action for the gravitational field.
In \cite{Perini2,Gaurav} it was shown that a nontrivial fixed point exists,
where the purely gravitational couplings are finite while those involving the scalar field vanish.
This is called a ``Gaussian matter fixed point'' (GMFP).
In the present paper we extend these results by considering an interacting scalar field
coupled to a class of higher derivative gravity theories which had been studied
previously in \cite{cpr,cpr2}.
We ask whether the scalar matter contribution is able to alter the results
of the purely gravitational part considerably.
In \cite{cpr2} the addition of minimally
coupled matter components to $R^2$-gravity (including all possible curvature invariants
up to quadratic order) showed that the nontrivial fixed point structure is maintained in
that case.
We will see that this is largely the case here too, but since we now consider
interacting scalars we will find that the dimension of the critical surface increases.

There are clearly many possible applications in cosmology.
Early work in this direction has been done in \cite{Reutercosmology},
using the beta functions of pure gravity.
Taking scalar fields into account could have significant renormalization group running effects in inflation. Without the necessity of asymptotic safety, in effective field theory calculations the beta functions derived here could be useful e.g. for inflation \cite{Weinberginflation}, or in models where the Higgs field is used as the inflaton field \cite{higgsflaton}.
Applications in the IR are possible,
for example along the lines of \cite{wetterichcosmology} or the much discussed modified theories of gravity with some
action based on different functional forms of the Ricci scalar (see e. g. \cite{fofr}).
We mention that the appearance of a scalar field in the low energy description of gravity 
has also been stressed in \cite{mottola}.
For a FRGE-based approach to that issue see also \cite{crehroberto}.

As in \cite{Gaurav} the analysis is based on a type of Wilsonian action $\Gamma_k$
called the ``effective average action" depending on an external energy scale $k$ which
can be formally defined by introducing an IR suppression in the functional integral
for the modes with momenta lower than $k$. This amounts to modifying the propagator,
leaving the interactions untouched. Then one can obtain a
Functional Renormalization Group Equation (FRGE) \cite{Wetterich} for the dependence of $\Gamma_k$ on $k$,
\begin{equation}
\label{eq:FRGE}
\partial_t \Gamma_k =
\frac{1}{2} {\rm STr} \Biggl[
\left(\frac{\delta^2\Gamma_k}{\delta \Phi \, \delta \Phi}+ \R _k\right)^{-1} \partial_t \R _k
\Biggr]
\end{equation}
where $t=\log(k/k_0)$.
$\Phi$ are all the fields present in the theory. STr is a generalized functional trace
including a minus sign for fermionic variables and a factor 2 for complex variables.
${\cal R}_k$ is the regulator that suppresses the contribution to the trace of fluctuations with momenta below $k$.
As the effective average action contains information about all the couplings in the theory, the FRGE
contains all the beta functions of the theory. In certain approximations one can use this equation to
reproduce the one-loop beta functions, but in principle the information one can extract from it is
nonperturbative, in the sense that it does not depend on the couplings being small.

A Quantum Field Theory (QFT) is asymptotically safe
if there exists a finite dimensional
space of action functionals (called the ultraviolet critical surface)
which in the continuum limit are attracted towards a Fixed Point (FP)
of the Renormalization Group (RG) flow.
For example, a free theory has vanishing beta functions, so it has a FP called the Gaussian FP.
Perturbation theory describes a neighbourhood of this point.
In a perturbatively renormalizable and asymptotically free QFT such as QCD,
the UV critical surface is parameterized by the couplings that
have positive or zero mass dimension.
Such couplings are called ``renormalizable'' or ``relevant''.
Asymptotic safety is a generalization of this behaviour
outside the perturbative domain. That means that the couplings could become
strong. The FRGE allows us to carry out calculations also in that regime.

Whether gravity is indeed asymptotically safe cannot yet be fully answered. However, since the
formulation of the FRGE by Wetterich \cite{Wetterich}, many results support this possibility in various
approximations  \cite{crehroberto, Reuter,Dou,LauscherR2,Codello,Machado:2007ea,Benedetti},
for reviews see \cite{NiedermaierReuter}. Since the first application to gravity the necessary tools have been
developed to make the approximation schemes more reliable including more couplings and studying their UV
behaviour. In the approximations so far used, gravity has a nontrivial fixed point with a finite dimensional UV
critical surface as is consistent with the requirements of asymptotic safety.

The most common approximation method is to expand the average effective action in derivatives and
to truncate the expansion at some order. In the case of scalar theory the lowest order of this expansion
is the local potential approximation (LPA), where one retains a standard kinetic term plus a generic
 potential. In the case of pure gravity, the derivative expansion involves operators that are powers of
curvatures and derivatives thereof. This has been studied systematically up to terms with four derivatives
in \cite{LauscherR2,Codello,Machado:2007ea,Benedetti} and for a limited class of operators
(namely powers of the scalar curvature) up to sixteen derivatives of the
metric \cite{cpr,cpr2}. In the case of scalar tensor theories of gravity, one
will have to expand both in derivatives of the metric and of the scalar field.

In this paper we will study the generalization of the action considered in
\cite{Perini1,Perini2,Gaurav} and \cite{cpr} of the form
\begin{equation}
\label{eq:action}
\Gamma_k [g,\phi ] = \int \, {d}^{d} x  \sqrt{g}
\left \{
F(\phi ^2, R) + \frac{1}{2} g^{\mu \nu} \nabla _{\mu} \phi \, \nabla_{\nu} \phi
\right \}
+S_{GF}+S_{gh} \, \ ,
\end{equation}
where $S_{GF}$ is a gauge fixing term to be specified below and $S_{gh}$ is
the corresponding ghost action. This action can be seen as a generalization
of the LPA where also terms with two or more derivatives of the metric are included.

This paper is organized in the following way. In section \ref{sectionFRGE} we will give the inverse
propagators resulting from the action (\ref{eq:action}) which have to be inserted into the FRGE to obtain the beta functions.
In section \ref{sectionGMFP} we describe the general properties of the GMFP. It is divided into two
sub-sections. In section \ref{sectionminmatterGMFP} we show that minimal
couplings are self-consistent in the sense that when matter couplings are
switched off then also their beta functions vanish.
In section \ref{sectionlinflowGMFP}, we analyze the linearized RG flow around the GMFP.
We find that the stability matrix has block diagonal form which allows us to
calculate its eigenvalues and eigenvectors in a recursive way.
In section \ref{sectionnonminscalar} we illustrate the existence of the GMFP and the properties
of the RG flow near the GMFP in specific truncations where scalar matter fields
are coupled nonminimally to gravity, including operators with up to four powers of
scalar curvature and quadratic in the scalar matter field.
In section \ref{sectionminscalar} we consider minimally coupled scalar tensor theory
including operators up to eight powers of scalar curvature and determine the
dimensionality of the UV critical surface.
We conclude in section \ref{sectionconclusions}.

\section{The FRGE for $F(\phi ^2, R) $ }
\label{sectionFRGE}

\subsection{Second variations}

Starting from the action given in eq. (\ref{eq:action}),
we expand $F(\phi ^2, R)$ in polynomial form in $\phi^2$ and $R$ as
\begin{equation}
\label{eq:F_exp}
F(\phi ^2, R) = V_0(\phi ^2) + V_1(\phi ^2) \, R + V_2(\phi ^2) \, R^2
+ V_3(\phi ^2) \, R^3 + \cdots + V_p(\phi ^2) \, R^p= \sum _{a=0}^p  V_a(\phi ^2) \, R^a \, \ .
\end{equation}
In order to evaluate the r.h.s. of eq. (\ref{eq:FRGE}) we calculate the second functional
derivatives of the functional given in eq. (\ref{eq:action}).
These can be obtained by expanding the action to second order in the quantum fields
around classical backgrounds $g_{\mu\nu}=\bar g_{\mu\nu}+h_{\mu\nu}$ and
$\phi=\bar\phi+\delta\phi$, where $\bar\phi$ is constant.
The gauge fixing action quadratic in $h_{\mu\nu}$ is chosen to be
\begin{equation}
\label{eq:GFaction}
S_{GF}=\frac{1}{2}\int d^dx\sqrt{\bar{g} }\,\chi_{\mu} \, {G}^{\mu\nu} \, \chi_{\nu} \, \ ,
\end{equation}
where $\chi_{\nu}=\bar{\nabla}^{\mu}h_{\mu\nu}-\frac{1+\rho}{d} \bar{\nabla}_{\nu}h^{\mu}_{\,\,\mu}$,
 $ G_{\mu\nu}=\bar{g}_{\mu\nu} \left( \alpha +\beta \bar{\Box} \right)$;
$\alpha$, $\beta$, and $\rho$ are the gauge parameters, we denote
$\bar{\Box} = \bar{\nabla}^{\mu}\bar{\nabla}_{\mu}$.

The gauge fixing action eq. (\ref{eq:GFaction}) gives rise to a ghost action consisting
of two parts, $S_{gh}=S_c+S_b$. The first part $S_c$ arises from the usual Fadeev-Popov prodedure
leading to the complex ghost fields $C_{\mu}$ and $\bar{C}_{\mu}$. It is given by
\begin{equation}
\label{eq:1st_ghost}
S_c = \int \, {d}^{d} x  \sqrt{\bar{g}} \bar{C}^{\mu} ( \alpha + \beta \bar{\Box} ) \left[ \delta ^{\nu}_{\, \,\mu} \bar{\Box}
+ \bar{R}^{\nu}_{\,\,\mu} + \frac{d-2-2 \rho}{d} \bar{\nabla} _{\mu} \bar{\nabla}^{\nu} \right] C_{\nu} \, \ .
\end{equation}
The second part $S_b$ arises for $\beta \neq 0$ and comes from the exponentiation of
a nontrivial determinant which requires the introduction
of real anti-commuting fields $b_{\mu}$ which are usually referred to as the
third ghost fields \cite{Buchbinder},
\begin{equation}
\label{eq:2nd_ghost}
S_b =  \frac{1}{2} \int \, {d}^{d} x  \sqrt{\bar{g}} \,  b_{\mu} \, G^{\mu\nu} \, b_{\nu} \, .
\end{equation}
These terms are already quadratic in the quantum fields.
Then the second variation of eq. (\ref{eq:action}) is given by
\begin{eqnarray}
\label{eq:2nd_var_Fphi2R}
\Gamma_k^{(2)}&=&
\frac{1}{2} \int \, {d}^{d} x  \sqrt{g}
\Biggl[ F(\phi ^2, R) \left \{ \frac{1}{4} h^2 - \frac{1}{2} h_{\mu \nu} h^{\mu \nu} \right \}
+ \frac{\partial F(\phi ^2, R)}{\partial R}
\Biggl \{ -h h^{\mu \nu} R_{\mu \nu}
- \frac{1}{2} h \Box h + \frac{1}{2} h^{\mu \nu} \Box h_{\mu \nu}
+ h^{\mu \alpha} h_{\alpha \beta} R_{\mu} ^{\beta}  \notag \\
&&  + h_{\mu \nu} R^{\mu \rho \nu \lambda} h_{\rho \lambda} - h_{\mu}^{\nu} \nabla ^{\mu} \nabla ^{\rho} h_{\rho \nu} + h \nabla ^{\mu} \nabla^{\nu} h_{\mu \nu}
\Biggr \}
+ \frac{\partial ^2 F(\phi ^2, R)}{\partial R^2}
\Biggl \{h^{\mu \nu} R_{\mu \nu}
\cdot h^{\alpha \beta} R_{\alpha \beta} - 2 h^{\mu \nu} R_{\mu \nu} \cdot
\nabla ^{\rho}\nabla^{\sigma} h_{\rho \sigma}  \nonumber \\
&&  + 2 h^{\mu \nu} R_{\mu \nu} \cdot \Box h + \nabla ^{\alpha} \nabla ^{\beta} h_{\alpha \beta} \cdot \nabla ^{\mu} \nabla ^{\nu} h_{\mu \nu } - 2 \Box h \cdot \nabla ^{\mu} \nabla ^{\nu} h_{\mu \nu } + \Box h \cdot \Box h
\Biggr \}
\Biggr] \notag \\
&& + \int \, {d}^{d} x  \sqrt{g}
\Biggl[
h \cdot \phi \frac{\partial F(\phi ^2, R)}{\partial \phi^2} \delta\phi
+ 2 \phi \, \delta\phi \frac{\partial ^2 F(\phi ^2, R) }{\partial R \,\partial \phi^2}
\Biggl \{ \nabla ^{\mu} \nabla ^{\nu} h_{\mu \nu}  - \Box h
- h^{\mu \nu} R_{\mu \nu} \Biggr \}
\Biggr] \nonumber \\
&& + \frac{1}{2} \int \, {d}^{d} x  \sqrt{g} \, \delta\phi \left[ - \Box + 2 \frac{\partial F(\phi ^2, R)}{\partial \phi^2} + 4 \phi ^2 \frac{\partial ^2 F(\phi ^2, R)}{\partial {(\phi^2)}^2} \right] \, \delta\phi\, + S_{GF} + S_{gh}
\, \ ,
\end{eqnarray}
where $\Box = \nabla^{\mu} \nabla_{\mu}$ and $h=h^{\mu}_{\mu}$.
Since we will never have to deal with the original metric $g_{\mu\nu}$ and scalar field $\phi$,
in order to simplify the notation, in the preceding
formula and everywhere else from now on we will remove the bars from the backgrounds.
As explained in \cite{Reuter}, the functional that obeys the FRGE  (\ref{eq:FRGE})
has a separate dependence on the background field $\bar g_{\mu\nu}$ and on a
``classical field'' $\left( g_{\mathrm{cl}} \right) _{\mu\nu}=\bar g_{\mu\nu}+ \left( h_{\mathrm{cl}} \right)_{\mu\nu}$,
where $ \left( h_{\mathrm{cl}} \right)_{\mu\nu}$ is the Legendre conjugate of the sources coupling linearly
to $ \left( h_{\mathrm{cl}} \right)_{\mu\nu}$. The same applies to the scalar field.
In this paper, like in most of the literature on the subject, we will restrict ourselves
to the case when $\left( g_{\mathrm{cl}} \right) _{\mu\nu}=\bar g_{\mu\nu}$ and $\phi_{\mathrm{cl}}=\bar\phi$.
From now on the notation $g_{\mu\nu}$ and $\phi$ will be used to denote equivalently
the ``classical fields'' or the background fields.

\subsection{Decomposition}

In order to simplify the terms and partially diagonalize
the kinetic operator, we perform a decomposition of $h_{\mu \nu}$ in tensor, vector, and scalar parts as in \cite{cpr,cpr2},
\begin{equation}
\label{eq:TTdecomp}
h_{\mu \nu} = h_{\mu \nu}^T + \nabla _{\mu} \xi _{\nu} + \nabla _{\nu} \xi _{\mu} + \nabla _{\mu} \nabla _{\nu} \sigma  - \frac{1}{d} g_{\mu \nu} \Box \sigma + \frac{1}{d} g_{\mu \nu} h
\end{equation}
where $h_{\mu \nu}^{T}$ is the (spin 2) transverse and traceless part,
$\xi_{\mu}$ is the (spin 1) transverse vector component,
$\sigma$ and $h$ are (spin 0) scalars.
This decomposition allows an exact inversion of the second variation under the restriction
to a spherical background. With that in mind, we work on a $d$-dimensional sphere.
For the spin-2 part, the inverse propagator is
\begin{equation}
\label{eq:sp2propgrav}
\frac{\delta^2\Gamma_k}{\delta h_{\mu \nu}^T\delta h_{\rho \sigma}^T}=
\Biggl[
\frac{1}{2} \frac{\partial F(\phi ^2, R)}{\partial R} \left \{ \Box  + \frac{2 (d-2) }{d (d-1) } R \right \} -\frac{1}{2} F(\phi ^2, R)
\Biggr] \, \delta^{\mu \nu,\rho \sigma} \, \ ,
\end{equation}
where $ \delta^{\mu \nu,\rho \sigma}
= \frac{1}{2} ( g^{\mu \rho} \, g^{\nu \sigma} + g^{\mu \sigma} \, g^{\nu \rho} )$.
For the spin-1 part it is
\begin{equation}
\label{eq:sp1propgrav}
\frac{\delta^2\Gamma_k}{\delta\xi_{\mu}\delta\xi_{\nu}}=
\left( \Box + \frac{R}{d} \right) \left[ ( \alpha + \beta \Box ) \left(\Box + \frac{R}{d} \right) - 2 \frac{\partial F(\phi ^2, R)}{\partial R} \, \frac{R}{d} + F(\phi ^2, R) \right]
\,g^{\mu \nu} \, .
\end{equation}
The two spin-0 components of the metric, $\sigma$ and $h$, mix with $\delta\phi$ resulting in an inverse propagator
given by a symmetric $3\times3$ matrix $S$ with the entries
\begin{gather}
S_{\sigma \sigma} = \left(1-\frac{1}{d} \right) (-\Box) \left(-\Box - \frac{R}{d-1} \right)
\Biggl[ \left(1-\frac{1}{d} \right) \left(-\Box - \frac{R}{d-1} \right) \left \{ \alpha + \beta \left( \Box + \frac{R}{d} \right) \right \} - \frac{1}{2} F(\phi ^2, R) \notag \\
  -\left( \frac{2-d}{2 d} \right) \left( -\Box - \frac{2R}{2 - d } \right) \frac{\partial F(\phi ^2, R)}{\partial R}
+ \left(1-\frac{1}{d} \right) (-\Box) \left(-\Box - \frac{R}{d-1} \right) \frac{\partial ^2 F(\phi ^2, R)}{\partial R^2}
 \biggr] \, \ , \notag \\
S_{\sigma h} = S_{h \sigma} =  \frac{1}{2} \left(1-\frac{1}{d} \right) (-\Box) \left(-\Box - \frac{R}{d-1} \right)
\Biggl[ \frac{2 \rho}{d} \left \{ \alpha + \beta \left( \Box + \frac{R}{d} \right) \right \} + \left( 1 - \frac{2}{d} \right) \frac{\partial F(\phi ^2, R)}{\partial R}\notag \\
+ 2 \left(1-\frac{1}{d} \right)  \left(-\Box - \frac{R}{d-1}\right) \frac{\partial ^2F(\phi ^2, R)}{\partial R^2}
\Biggr] \, \ , \qquad
S_{\sigma \phi} = S_{\phi \sigma } = 2 \phi  \left(1-\frac{1}{d} \right) (-\Box) \left(-\Box - \frac{R}{d-1} \right) \frac{\partial ^2F(\phi ^2, R)}{\partial R \, \partial \phi^2}
\, \ , \notag \\
S_{h h} = \Biggl[ -\Box \left \{ \alpha + \beta \left( \Box + \frac{R}{d} \right) \right \} \left( \frac{\rho}{d} \right)^2 + \frac{d - 2}{4 d} F(\phi ^2, R)
+ \left(1-\frac{1}{d} \right) \left(\frac{1}{2} - \frac{1}{d} \right) \left(-\Box - \frac{2R}{d-1} \right)  \frac{\partial F(\phi ^2, R)}{\partial R} \notag \\
+ \left(1-\frac{1}{d} \right)^2 \left(-\Box - \frac{R}{d-1} \right)^2 \frac{\partial ^2F(\phi ^2, R)}{\partial R^2}
\Biggr] \, \ , \qquad
S_{h \phi} = S_{\phi h} = \phi \frac{\partial F(\phi ^2, R)}{\partial \phi^2}
+ 2 \phi \left(1-\frac{1}{d} \right) \left(-\Box - \frac{R}{d-1} \right)
\frac{\partial ^2 F(\phi ^2, R)}{\partial R \, \partial \phi^2} \, \ , \notag \\
S_{\phi \phi} = -\Box + 2 \frac{\partial F(\phi ^2, R)}{\partial \phi^2}
 + 4 \phi ^2 \frac{\partial ^2 F(\phi ^2, R)}{\partial (\phi^2)^2 } \, \ .
\label{spin0propgrav}\end{gather}
As discussed in more detail in \cite{cpr2}, to match the trace-spectra of the
Laplace-operator acting on $h_{\mu\nu}$ with those obtained for the constrained
fields after the decomposition, the first eigenmode of the operator trace over the vector
contribution and the first two eigenmodes of the operator trace over the $\sigma$ contribution
have to be omitted. The trace over the $h$ and $\delta\phi$ components should be taken
over the whole operator spectrum instead. To handle the mixing of the scalar components
in an easy way, we subtract first the two first eigenmodes from the complete scalar contribution
from the matrix $S$ and then add the first two trace modes which should have been retained
for $h$ and $\delta \phi$. This requires to take into account a further scalar matrix $B$ formed
by the components of $h$, $\phi$ and their mixing term. It is given by
\begin{equation}
\label{eq:sp0_propgrav2}
B=\left(
\begin{array}{cc}
S_{hh}& S_{h\phi}\\
S_{h\phi}& S_{\phi\phi}
\end{array}
\right) \, \ ,
\end{equation}
whose trace contribution to the FRGE will be
calculated on the first two eigenmodes of the spectrum of the Laplacian.

Again, in order to diagonalize the kinetic operators occurring in the ghost actions eqs. (\ref{eq:1st_ghost})
and (\ref{eq:2nd_ghost}), we perform a decomposition of the ghost fields $C_{\mu}$, $\bar{C}_{\mu}$ and  $b_{\mu}$
into transverse and longitudinal parts,
\begin{gather}
\label{eq:ghbreak}
\bar{C}^{\mu} = \bar{C}^{\mu T} + \nabla ^{\mu} \bar{C}, \hspace{10mm} C_{\mu} = C^{T}_{\mu} + \nabla_{\mu} C, \hspace{10mm}  b_{\mu} = b^{T}_{\mu} + \nabla _{\mu} b,
\end{gather}
with
$\nabla_{\mu} \bar{C}^{\mu T} = 0$,
$\nabla^{\mu} C^{T}_{\mu} =0$ and
$\nabla^{\mu} b^{T}_{\mu} =0$.

After this decomposition, the inverse propagators for the vector and scalar components of the ghost and third ghost fields are
\begin{eqnarray}
\label{eq:1st_gh_sp1_prop}
\frac{\delta^2\Gamma_k}{\delta {\bar C}^T_{\mu} \, \delta C^T_{\nu} }&=&
(\alpha +\beta \Box ) \left( \Box+ \frac{R}{d} \right) \, g^{\mu \nu} \, \ , \\
\label{eq:1st_gh_sp0_prop}
\frac{\delta^2\Gamma_k}{\delta \bar{C} \, \delta C}&=&
\frac{2( d - 1 - \rho) }{d} (-\Box) \left[ \alpha + \beta \left( \Box + \frac{R}{d} \right) \right] \left[ \Box + \frac{R}{ d - 1 - \rho } \right] \, \ ,\\
\label{eq:2nd_gh_sp1_prop}
\frac{\delta^2\Gamma_k}{\delta b^T_{\mu} \,  \delta b^T_{\nu} }&=&(\alpha +\beta \Box ) \, g^{\mu \nu} \, \ , \\
\label{eq:2nd_gh_sp0_prop}
\frac{\delta^2\Gamma_k}{\delta b \, \delta b}&=& -\Box \left[ \alpha + \beta \left( \Box + \frac{R}{d} \right) \right] \, .
\end{eqnarray}

\subsection{Contributions by Jacobians}

The decomposition of $h_{\mu \nu}$, $\bar C_{\mu}$, $C_{\mu}$, and $b_{\mu}$ gives
rise to nontrivial Jacobians in the path integral, given by
\begin{gather}
\label{eq:jacob_grav}
J_{\xi} = \left[ det' \left( -\Box - \frac{R}{d} \right) \right]^{1/2} \, \ , \ \
J_{\sigma}= \left[ det'' \left \{\Box \left( \Box + \frac{R}{d-1} \right) \right \} \right]^{1/2} \, \ , \ \
J_c=[ det ' (- \Box ) ]^{-1} \, \ , \ \
J_b=[ det ' (- \Box ) ]^{-1} \, \ .
\end{gather}
These Jacobians can be absorbed by field redefinitions which however
introduce terms which involve noninteger powers of the Laplacian. To avoid technical difficulties, we therefore prefer
to exponentiate these Jacobians by the introduction of auxiliary
anticommuting and commuting fields according to the sign of the exponent
of the determinant, see also \cite{cpr,cpr2}. One has to take their contribution into account while writing the FRGE.

\section{The Gaussian Matter fixed point}

\label{sectionGMFP}

The running of $V_a(\phi^2)$ is calculated from the FRGE as
\begin{equation}
\label{eq:run_F}
(\partial _t V_a)[\phi^2] = \frac{1}{\rm Vol} \, \frac{1}{a!} \, \frac{\partial ^a (\partial _t \Gamma_k)[ \phi^2, R] }{\partial R^a}
\end{equation}
where $(\partial _t \Gamma_k)[ \phi^2, R]$ is obtained for various fields in an analogous way as in \cite{cpr,cpr2}.
Rescaling all fields with respect to the cutoff scale $k$, we obtain the dimensionless quantities
$\tilde\phi=k^{\frac{2-d}{2}}\phi$, $\tilde R=k^{-2}R$ and
 $\tilde{V}_a(\tilde{\phi}^2) = k^{-(d-2a) } V_a(\phi^2)$.
These dimensionless quantities we can use to analyze the RG flow and its FP structure. From the running of $ V_a(\phi^2)$ one
can calculate the running of
$\tilde V_a(\tilde{\phi}^2)$ using
 \begin{eqnarray}
 \label{eq:dlessF_beta}
(\partial _t \tilde{V}_a)[ \tilde{\phi}^2] &=& -(d-2a) \tilde{V}_a( \tilde{\phi}^2 )
+ (d-2) \tilde{\phi} ^2 \, \tilde{V}_a^{\prime} (\tilde{\phi} ^2)
+ k^{-(d-2a) } \, (\partial _t V_a)[\phi^2] \,
\end{eqnarray}
where the last term is calculated using eq. (\ref{eq:run_F}).
A FP is a solution of the infinite set of functional equations
$
\partial_t\tilde V_{a}=0
$
for $a=0,\ldots ,\infty$.
This means that, at the FP, for each $a$ the function $\tilde V_{a}(\tilde \phi^2)$
is $k$-independent, or equivalently that each coefficient of its Taylor
expansion is $k$-independent.
Since we assume that each $\tilde V_{a}$ is analytic it can be Taylor expanded around $\tilde\phi^2=0$,
and therefore
\begin{equation}
\label{fpdef}
\partial_t\tilde V_{a}^{(i)}(0)=0
\end{equation}
for $i=0, \ldots ,\infty$, where the superscript $i$ denotes the $i$-th derivative with respect to
$\tilde\phi^2$.

\subsection{Minimal matter coupling of gravity at the GMFP}
\label{sectionminmatterGMFP}
The existence of a Gaussian Matter Fixed Point (GMFP), where all the matter couplings approach zero for $k \to \infty$
and only the purely gravitational couplings have nontrivial values,
was observed for finite polynomial truncations in \cite{Perini2}.
In \cite{Gaurav}, its existence was proven for effective average actions of the form
\begin{equation}
\label{eq:act_part1}
\Gamma_k [g,\phi ] = \int\,{d}^{d} x \, \sqrt{g} \left(V_0(\phi^2)+V_1(\phi^2) \, R
+\frac{1}{2} \,  g^{\mu \nu} \, \partial_{\mu} \phi \partial_{\nu} \phi \right)
+S_{GF}+S_{gh}.
\end{equation}
The existence of a GMFP can be shown to hold for the more general class of effective average actions considered in this paper.
By definition, a GMFP is a point where $\tilde V_{a}$ are $\tilde\phi^2$-independent, {\it i.e.}
\begin{equation}
\label{gmfpdef}
\tilde V_{a}^{(i)}(0)=0
\end{equation}
for $i=1,\ldots ,\infty$.
In this subsection we will prove that with the ansatz in eq. (\ref{gmfpdef})
all the equations in (\ref{fpdef}) with $i=1,\ldots ,\infty$ are identically satisfied,
thus leaving only the equations with $i=0$ to be solved.
We will give numerical solutions of these remaining equations
for $a=0,1\ldots ,8$ in section \ref{numericalresults}.

Now we explicitly analyze the structure of $\partial _t F$ related to the second variation of the
effective average action given in eq. (\ref{eq:action}) for the various field components.
The second variation for $h_{\mu \nu}^T$ and $\xi_{\mu}$ has the form
\begin{equation}
\label{eq:sp2sp1_form}
\left. \Gamma^{(2)}_k \right |_{T,V} = f(z, \, R)+f_a ( z, \, R) \, V_a  \, \ ,
\end{equation}
where we denote $z:=-\Box$. The functional form for $\left. \Gamma^{(2)}_k \right |_{T,V}$
is motivated by eqs. (\ref{eq:sp2propgrav}) and (\ref{eq:sp1propgrav}) from which we notice that
it depends on $V_a$ at most linearly, with coefficients being functions of $z$ and $R$,
which are denoted here by $f(z, \, R)$ and $f_a ( z, \, R)$.

For the scalar part, the second variation has the form
\begin{gather}
\label{eq:Gamma2_s}
\left. \Gamma^{(2)}_k \right |_s =
\left(
\begin{array}{c c c}
l^{11}(z, R)+f^{11}_a (z, R) \, V_a           &   l^{12}(z, R)+f^{12}_a(z, R) \, V_a               &   g^1_a(z, R) \, \phi \, V_a^{\prime}   \\
l^{12}(z, R) +f^{12}_a(z, R) \, V_a           &    l^{22}(z, R)   +f^{22}_a(z, R) \, V_a           &   g^2_a(z, R) \, \phi \, V_a^{\prime}\\
g^1_a(z, R) \, \phi \, V_a^{\prime}  &  g^2_a(z, R) \, \phi \, V_a^{\prime}  &   z + R^a ( 2 \, V_a^{\prime} + 4 \, \phi^2 \, V_a^{\prime \prime} ) \\
\end{array}
\right) \, \ ,
\end{gather}
where a prime denotes derivative with respect to $\phi^2$.
Again the functional form for $\left. \Gamma^{(2)}_k\right|_s$ is motivated by eq. (\ref{spin0propgrav})
which clearly tells that entries $S_{\sigma \sigma}$, $S_{\sigma h}$, and $S_{hh}$ depend at most linearly on
$V_a$, while the entries $S_{\phi \sigma}$ and $S_{\phi h}$ are linear combinations of $\phi \, V_a^{\prime}$.
The coefficients in these linear combinations are functions of $z$ and $R$
denoted here by $l^{ij}(z,R)$ and $g_a^i(z,R)$.

For the ghost part the second variation has the form
\begin{equation}
\label{eq:Gamma2_gh}
\left. \Gamma^{(2)}_k \right |_{gh} = D(z, \, R) \, \ .
\end{equation}
This can be verified from eqs. (\ref{eq:1st_gh_sp1_prop},
\ref{eq:1st_gh_sp0_prop}, \ref{eq:2nd_gh_sp1_prop} and \ref{eq:2nd_gh_sp0_prop}).
We first consider the contributions from $h_{\mu \nu}^T$ and $\xi_{\mu}$.
Since for them the second variation has the form given by
eq. (\ref{eq:sp2sp1_form}), the modified inverse propagator
$\mathcal{P}_k:=\Gamma_k^{(2)}+{\mathcal R}_k$ and the cutoff ${\mathcal R}_k$
will have the functional form
\begin{equation}
\label{eq:PR_TV}
\mathcal{P}_k = f( P_k, R)+ f_a( P_k , R) V_a  \, \ , \qquad
\mathcal{R}_k = f(P_k, R) - f(z, R)+\{ f_a(P_k,R) - f_a( z , R) \} V_a \, \ ,
\end{equation}
where we have simply replaced $z$  by
$P_k(z):=z+R_k(z)$ to obtain the modified inverse propagator.
$R_k(z)$ is a profile function which tends to $k^2$ for $z\rightarrow 0$ and approaches zero rapidly for $z>k^2$.
The RG-time derivative of the cutoff ${\mathcal R}_k$ in eq. (\ref{eq:PR_TV}) is
\begin{equation}
\label{eq:tder_R_TV}
\partial _t \mathcal{R}_k = \partial _t f(P_k, R)+ \partial _t f_a(P_k, R) \, V_a +  \{ f_a(P_k,R) - f_a( z , R) \} \, \partial _t V_a \, \ .
\end{equation}
Using eq. (\ref{eq:tder_R_TV}) in the FRGE one finds that the contributions from $h_{\mu \nu}^T$ and
$\xi_{\mu}$ have the form
\begin{equation}
\label{eq:tder_Va_TV}
\partial _t V_a = H_a (V_c) + H_{ab}(V_c) \partial _t V_b \, \ .
\end{equation}
This can be justified by noticing that $\partial _t \mathcal{R}_k$ given by eq. (\ref{eq:tder_R_TV})
depends at most linearly on $\partial _t V_b$. On the r.h.s. of the
FRGE, $\partial _t \mathcal{R}_k$ occurs in the numerator, while the
denominator contains the modified inverse propagator
given in eq. (\ref{eq:PR_TV}) which depends at most linearly on $V_a$. So we
find that the r.h.s of the FRGE depends at most linearly on $\partial _t V_a$.
The coefficients in front of  $\partial _t V_a$ are
functionals of $V_a$ and are denoted by $H_a(V_c)$ and $H_{ab}(V_c)$.

The contributions from the ghost parts will be simpler. Since they do not depend on the potentials,
they will only give a constant contribution to $H_a$. The contributions from
the scalars are more involved due to the matrix structure.
The modified inverse scalar propagator is obtained by replacing all $z$ with $P_k$ in
eq. (\ref{eq:Gamma2_s}).

The cutoff is constructed in the usual way by subtracting the inverse propagator
from the modified inverse propagator. This cutoff can be written as
\begin{eqnarray}
\label{eq:cutoff_s}
\mathcal{R}_k^s &=&
\left(
\begin{array}{c c c}
l^{11}(P_k, R) - l^{11}(z, R)      &      l^{12}(P_k, R) - l^{12}(z, R)    &    0 \\
 l^{12}(P_k, R) - l^{12}(z, R)     &      l^{22}(P_k, R) - l^{22}(z, R)     &    0 \\
 0                                           &                   0                                &   P_k -z \\
\end{array}
\right) \notag \\
&& + \left(
\begin{array}{c c c}
f_a^{11}(P_k, R) - f_a^{11}(z, R)      &      f_a^{12}(P_k, R) - f_a^{12}(z, R)    &    0 \\
 f_a^{12}(P_k, R) - f_a^{12}(z, R)     &      f_a^{22}(P_k, R) - f_a^{22}(z, R)     &    0 \\
 0                                           &                   0                                &   0 \\
\end{array}
\right) \, V_a \notag \\
&& + \left(
\begin{array}{c c c}
0                                         & 0                                         & g^1_a(P_k, R) - g^1_a(z, R) \\
0                                         & 0                                         & g^2_a(P_k, R) - g^2_a(z, R) \\
g^1_a(P_k, R) - g^1_a(z, R) & g^2_a(P_k, R) - g^2_a(z, R) & 0 \\
\end{array}
\right) \, \phi V_a^{\prime} \, \ .
\end{eqnarray}
Then the $t$ derivative of the cutoff given in eq. (\ref{eq:cutoff_s}) is
\begin{eqnarray}
\label{eq:tder_cutoff_s}
\partial _t \mathcal{R} _k^s &=&
\left(
\begin{array}{c c c}
\partial _t l^{11} (P_k, R)  &   \partial_t l^{12}(P_k, R)    &   0  \\
\partial _t l^{12} (P_k, R)  &   \partial_t l^{22}(P_k, R)    &   0  \\
              0                        &                  0                       &  \partial _t P_k \\
\end{array}
\right)
+ \left(
\begin{array}{c c c}
\partial _t f_a^{11} (P_k, R)  &   \partial_t f_a^{12}(P_k, R)    &   0  \\
\partial _t f_a^{12} (P_k, R)  &   \partial_t f_a^{22}(P_k, R)    &   0  \\
              0                        &                  0                              &    0 \\
\end{array}
\right) \, V_a  \notag \\
&& +
\left(
\begin{array}{c c c}
f_a^{11}(P_k, R) - f_a^{11}(z, R)      &      f_a^{12}(P_k, R) - f_a^{12}(z, R)    &    0 \\
 f_a^{12}(P_k, R) - f_a^{12}(z, R)     &      f_a^{22}(P_k, R) - f_a^{22}(z, R)     &    0 \\
 0                                           &                   0                                &   0 \\
\end{array}
\right) \, \partial _t V_a
+
\left(
\begin{array}{c c c}
0                                     &   0                                     &   \partial _t g^1_a( P_k ,R) \\
0                                     &    0                                    &    \partial _t g^2_a( P_k ,R)  \\
\partial _t g^1_a( P_k ,R) & \partial _t g^2_a( P_k ,R)  & 0 \\
\end{array}
\right) \, \phi V_a^{\prime} \notag \\
&& + \left(
\begin{array}{c c c}
0                                         & 0                                         & g^1_a(P_k, R) - g^1_a(z, R) \\
0                                         & 0                                         & g^2_a(P_k, R) - g^2_a(z, R) \\
g^1_a(P_k, R) - g^1_a(z, R) & g^2_a(P_k, R) - g^2_a(z, R) & 0 \\
\end{array}
\right) \, \phi \partial _t V_a^{\prime} \, \ .
\end{eqnarray}

The modified propagator for scalars is the matrix inverse of
eq. (\ref{eq:Gamma2_s}) with $z$ replaced by $P_k$.
It is given by
\begin{equation}
\label{eq:matinv_P}
\left( \mathcal{P}_k^s \right)^{-1}  = \frac{1}{Det \mathcal{P}_k^s } \, Adj \left( \mathcal{P}_k^s \right) \, \ .
\end{equation}
where $Adj \left( \mathcal{P}_k^s \right)$ denotes the adjoint of the
matrix $\left( \mathcal{P}_k^s \right)$ (the matrix of cofactors).
The determinant is a functional depending only on
$V_a$, $\phi^2 \, V_a^{\prime} V_b^{\prime}$, and $2 \, V_a^{\prime}
 + 4 \phi^2 \, V_a^{\prime \prime}$.
This can be easily derived from the modified inverse propagator obtained from
eq. (\ref{eq:Gamma2_s}).

All entries of the adjoint of $\mathcal{P}_k^s$ consist of cofactors, thus it has the form
\begin{equation}
\label{eq:Adj_P_s}
Adj \left( \mathcal{P}_k^s \right) =
\left(
\begin{array}{c c c}
A^{11}\left( V_a, \phi^2 \, V_a^{\prime} V_b^{\prime}, 2 \, V_a^{\prime} + 4 \phi^2 \, V_a^{\prime \prime} \right) &
A^{12}\left( V_a, \phi^2 \, V_a^{\prime} V_b^{\prime}, 2 \, V_a^{\prime} + 4 \phi^2 \, V_a^{\prime \prime} \right) &
A^{13} (V_a) \phi \, V_a^{\prime}\\
A^{21}\left( V_a, \phi^2 \, V_a^{\prime} V_b^{\prime}, 2 \, V_a^{\prime} + 4 \phi^2 \, V_a^{\prime \prime} \right) &
A^{22} \left( V_a, \phi^2 \, V_a^{\prime} V_b^{\prime}, 2 \, V_a^{\prime} + 4 \phi^2 \, V_a^{\prime \prime} \right) &
A^{23} (V_a) \phi \, V_a^{\prime}\\
A^{31} (V_a) \phi \, V_a^{\prime} &
A^{32} (V_a) \phi \, V_a^{\prime} &
A^{33} (V_a)\\
\end{array}
\right) \, \ ,
\end{equation}
where each entry depends additionally on $P_k$ and $R$. In order to calculate the RG trace, we
multiply $\left( \mathcal{P}_k^s \right)^{-1} $ with $\partial _t \mathcal{R} _k^s$ and then take the matrix trace. Doing this
we note that $\phi V_a^{\prime}$ is either multiplied with another $\phi V_a^{\prime}$ or it is multiplied with
$\phi \partial _t V_a^{\prime}$.
So the scalar contribution to the FRGE has the form
\begin{eqnarray}
\label{eq:Va_beta_s}
\left. \partial _t V_a \right |_s  &=&
H_a\left( V_a, \phi^2 \, V_a^{\prime} V_b^{\prime}, 2 \, V_a^{\prime} + 4 \phi^2 \, V_a^{\prime \prime} \right) +
H_{ab}\left( V_a, \phi^2 \, V_a^{\prime} V_b^{\prime}, 2 \, V_a^{\prime} + 4 \phi^2 \, V_a^{\prime \prime} \right) \,
\partial _t V_b \notag \\
&& + H_{abc}\left( V_a, \phi^2 \, V_a^{\prime} V_b^{\prime}, 2 \, V_a^{\prime} + 4 \phi^2 \, V_a^{\prime \prime} \right)
\, \phi^2 \, V_b^{\prime} \, \partial _t V_c^{\prime} \, .
\end{eqnarray}
The contributions from the transverse traceless tensor and transverse vector
can also be combined in the  above expression to write the full FRGE
contribution in the same way as above. Then $\partial _t F = R^a \, \partial _t V_a$.

After having calculated the structural form for the running of $V_a(\phi ^2)$,
we use it to calculate the dimensionless beta functional
using eq. (\ref{eq:dlessF_beta}), which gives
\begin{eqnarray}
\label{eq:dless_Va_beta}
(\partial _t \tilde{V}_a)[\tilde{\phi}^2] &=& -(d-2 a) \tilde{V}_a  + (d-2) \tilde{\phi}^2 \tilde{V}_a^{\prime}
+\tilde{H}_a \left( \tilde{V}_a, \, \tilde{\phi}^2 \, \tilde{V}_a^{\prime} \, \tilde{V}_b^{\prime},
\, 2 \tilde{V}_a^{\prime} + 4 \tilde{\phi}^2 \, \tilde{V}_a^{\prime \prime } \right) \notag \\
&& + \tilde{H}_{ab} \left( \tilde{V}_a, \, \tilde{\phi}^2 \, \tilde{V}_a^{\prime} \, \tilde{V}_b^{\prime},
\, 2 \tilde{V}_a^{\prime} + 4 \tilde{\phi}^2 \, \tilde{V}_a^{\prime \prime } \right) \,
\left \{ (d-2 b) \tilde{V}_b  - (d-2) \tilde{\phi}^2 \tilde{V}_b^{\prime} + ( \partial _t \tilde{V}_b ) [\tilde{\phi}^2 ] \right \} \notag \\
&& + \tilde{H}_{abc} \left( \tilde{V}_a, \, \tilde{\phi}^2 \, \tilde{V}_a^{\prime} \, \tilde{V}_b^{\prime},
\, 2 \tilde{V}_a^{\prime} + 4 \tilde{\phi}^2 \, \tilde{V}_a^{\prime \prime } \right) \, \tilde{\phi}^2 \tilde{V}_b^{\prime}
\left \{ (d-2 c) \tilde{V}_c^{\prime}  - (d-2) \, \left(
\tilde{\phi}^2 \tilde{V}_c^{\prime \prime } + \tilde{V}_c^{\prime }
\right) + ( \partial _t \tilde{V}_c )^{\prime} [\tilde{\phi}^2 ] \right \} \, \ .
\end{eqnarray}
Inserting eq. (\ref{fpdef})  in eq. (\ref{eq:dless_Va_beta}) we get the fixed point equation
\begin{eqnarray}
\label{eq:FPeq_Va}
0 &=& -(d-2 a) \tilde{V}_a  + (d-2) \tilde{\phi}^2 \tilde{V}_a^{\prime}
+\tilde{H}_a \left( \tilde{V}_a, \, \tilde{\phi}^2 \, \tilde{V}_a^{\prime} \, \tilde{V}_b^{\prime},
\, 2 \tilde{V}_a^{\prime} + 4 \tilde{\phi}^2 \, \tilde{V}_a^{\prime \prime } \right) \notag \\
&& + \tilde{H}_{ab} \left( \tilde{V}_a, \, \tilde{\phi}^2 \, \tilde{V}_a^{\prime} \, \tilde{V}_b^{\prime},
\, 2 \tilde{V}_a^{\prime} + 4 \tilde{\phi}^2 \, \tilde{V}_a^{\prime \prime } \right) \,
\left \{ (d-2 b) \tilde{V}_b  - (d-2) \tilde{\phi}^2 \tilde{V}_b^{\prime} \right \} \notag \\
&& + \tilde{H}_{abc} \left( \tilde{V}_a, \, \tilde{\phi}^2 \, \tilde{V}_a^{\prime} \, \tilde{V}_b^{\prime},
\, 2 \tilde{V}_a^{\prime} + 4 \tilde{\phi}^2 \, \tilde{V}_a^{\prime \prime } \right) \, \tilde{\phi}^2 \tilde{V}_b^{\prime}
\left \{ (d-2 c) \tilde{V}_c^{\prime}  - (d-2) \, \left(
\tilde{\phi}^2 \tilde{V}_c^{\prime \prime } + \tilde{V}_c^{\prime }
\right)  \right \} \, \ .
\end{eqnarray}
The above equation is identically satisfied when we take its
Taylor expansion around $\tilde{\phi}^2=0$ and use eq. (\ref{gmfpdef}).
For example, taking one derivative with respect to $\tilde\phi^2$ gives
\begin{eqnarray}
\label{eq:FPeq_Va_1der}
0 &=& -(d-2 a) \tilde{V}_a^{\prime}  + (d-2) \left\{ \tilde{\phi}^2 \tilde{V}_a^{\prime \prime } + \tilde{V}_a^{\prime} \right \}
 + \frac{\delta \tilde H_a}{\delta \tilde V_c} \tilde {V}^{\prime}_c
+ \frac{\delta \tilde H_a}{\delta (\tilde{\phi}^2 \tilde V^{\prime}_c \tilde V^{\prime}_d )  } (  \tilde V^{\prime}_c \tilde V^{\prime}_d
+ \tilde{\phi}^2  \tilde V^{\prime \prime}_c \tilde V^{\prime}_d +\tilde{\phi}^2  \tilde V^{\prime \prime}_d \tilde V^{\prime}_c )
\notag \\
&& + \frac{\delta \tilde H_a}{\delta(2 \tilde V^{\prime}_c  + 4 \tilde{\phi}^2 \tilde V^{\prime \prime}_c)} (  2 \tilde V^{(2)}_c  + 4 \tilde V^{(2)}_c + 4 \tilde{\phi}^2 \tilde V^{(3)}_c )
+  \left \{ (d-2 b) \tilde{V}_b^{\prime}  - (d-2) \tilde{\phi}^2 \tilde{V}_b^{\prime \prime } -(d-2) \tilde{V}_b^{\prime} \right \} \,
\tilde H_{ab} \notag \\
&& + \left \{ (d-2 b) \tilde{V}_b  - (d-2) \tilde{\phi}^2 \tilde{V}_b^{\prime }\right \}
\Biggl(
\frac{\delta \tilde H_{ab} }{\delta \tilde V_c} \, \tilde {V}^{\prime}_c
 + \frac{\delta \tilde H_{ab} }{\delta (\tilde{\phi}^2 \, \tilde V^{\prime}_c \, \tilde V^{\prime}_d )  } \,
 (  \tilde V^{\prime}_c \, \tilde V^{\prime}_d + \tilde{\phi}^2  \, \tilde V^{\prime \prime}_c \, \tilde V^{\prime}_d
 +\tilde{\phi}^2  \, \tilde V^{\prime \prime}_d \, \tilde V^{\prime}_c ) \notag \\
&& + \frac{\delta \tilde H_{ab} }{\delta(2 \tilde V^{\prime}_c  + 4 \tilde{\phi}^2 \tilde V^{\prime \prime}_c)} \, (  2 \tilde V^{(2)}_c
 + 4 \tilde V^{(2)}_c + 4 \tilde{\phi}^2 \tilde V^{(3)}_c )
 \Biggr)
 + \tilde{V}_b^{\prime} \, \left \{ (d-2 c) \tilde{V}_c  - (d-2) \tilde{\phi}^2 \tilde{V}_c^{\prime }\right \} \, \tilde H_{abc}
 \notag \\
 && + \tilde{\phi}^2 \tilde{V}_b^{\prime \prime } \, \left \{ (d-2 c) \tilde{V}_c  - (d-2) \tilde{\phi}^2 \tilde{V}_c^{\prime }\right \} \, \tilde H_{abc}
 + \tilde{\phi}^2 \tilde{V}_b^{\prime } \, \left \{ (d-2 c) \tilde{V}_c^{\prime }  - (d-2) \tilde{\phi}^2 \tilde{V}_c^{\prime\prime }
 -(d-2) \tilde{V}_c^{\prime } \right \} \, \tilde H_{abc} \notag \\
&& + \tilde{\phi}^2 \tilde{V}_b^{\prime } \, \left \{ (d-2 c) \tilde{V}_c  - (d-2) \tilde{\phi}^2 \tilde{V}_c^{\prime }\right \}
\Biggl(
\frac{\delta \tilde H_{abc} }{\delta \tilde V_d} \, \tilde {V}^{\prime}_d
 + \frac{\delta \tilde H_{abc} }{\delta (\tilde{\phi}^2 \, \tilde V^{\prime}_d \, \tilde V^{\prime}_e )  } \,
 (  \tilde V^{\prime}_d \, \tilde V^{\prime}_e + \tilde{\phi}^2  \, \tilde V^{\prime \prime}_d \, \tilde V^{\prime}_e
 +\tilde{\phi}^2  \, \tilde V^{\prime \prime}_e \, \tilde V^{\prime}_d ) \notag \\
&& + \frac{\delta \tilde H_{abc} }{\delta(2 \tilde V^{\prime}_d  + 4 \tilde{\phi}^2 \tilde V^{\prime \prime}_d)} \, (  2 \tilde V^{(2)}_d
 + 4 \tilde V^{(2)}_d + 4 \tilde{\phi}^2 \tilde V^{(3)}_d )
 \Biggr) \, \ .
\end{eqnarray}
Setting $\tilde{\phi}^2 =0$ and using the GMFP conditions, we see that
the right hand side will be zero.
One can take successive derivatives to verify that this property indeed holds when higher derivatives are taken.
The only equation which is not automatically solved in this way is the one
where we evaluate eq. (\ref{eq:FPeq_Va}) at $\tilde{\phi}^2 =0$ and use eq. (\ref{gmfpdef}). This is just the FP equation
for an $f(R)$ theory with a single minimally coupled scalar.
We will solve these equations in section \ref{numericalresults}.

\subsection{Linearized Flow around the GMFP}
\label{sectionlinflowGMFP}

The attractivity properties of a FP are determined by the signs of the
critical exponents defined to be minus the eigenvalues of the
linearized flow matrix, the so-called stability matrix, at the FP.
The eigenvectors corresponding to negative eigenvalues (positive critical
exponent) span the UV critical surface.
At the Gaussian FP the critical exponents are equal to the mass dimension of each coupling,
so the relevant couplings are the ones that are
power--counting renormalizable (or marginally renormalizable).
In a pertubatively renormalizable theory they are usually finite in number.

At the GMFP, the situation is more complicated as the eigenvalues being
negative or positive do not correspond to couplings being relevant or
irrelevant. In principle, at the GMFP the eigenvectors corresponding to negative
eigenvalues get contributions from all the couplings present in the truncation,
thus making it more difficult to find the fixed point action. Thus understanding
the properties of the stability matrix around the GMFP becomes crucial.

Therefore we now discuss the structure of the linearized flow around the GMFP.
It is convenient to Taylor expand
the potentials $V_a(\phi^2)$ as
\begin{equation}
\label{eq:Va_exp}
V_a(\phi ^2) = \sum _{i=0}^q \lambda ^{(a)}_{2 i }(k) \phi ^{2 i },
\end{equation}
where $\lambda^{(a)}_{2i}$ are the corresponding couplings with mass dimension $d-2a-i(d-2)$.
We are assuming a finite truncation with up to $p$ powers of $R$, {\it
i.e.} $a$ going from $0$ to $p$, and $q$ powers of $\phi^2$.
In practice it has been possible to deal with $p\leq8$;
as we shall see, it is possible to understand the structure of the theory
for any polynomial in $\phi^2$, so one could also let $q\to\infty$.
Rescaling these couplings with respect to the RG scale defines dimensionless couplings
$\tilde{\lambda} ^{(a)}_{2 i } = k^{d-2a-i(d-2)}\lambda ^{(i)}_{2 i }$
and the corresponding beta functions $\beta_{2i}^{(a)}=\partial_t \lambda_{2i}^{(a)}$.

The stability matrix is defined as
\begin{equation}
\label{eq:stab_mat}
\left( M_{ij} \right) _{ab} =
\left. \frac{ \delta \left ( \frac{1}{i!} \partial _t \tilde{V}_a^{(i)}(0) \right) }{ \delta  \left ( \frac{1}{j!} \tilde{V}_b^{(j)}(0)  \right)}
\right|_{ FP}
=\left.\frac{\partial \beta_{2i}^a}{\partial\tilde\lambda_{2j}^{(b)}}\right|_{ FP}
\end{equation}
Using the above definitions, numerical results tell that the stability matrix $M$ has the form
\begin{equation}
\label{eq:stmatform}
\left(
\begin{array}{ccccc}
M_{00} & M_{01} & 0 & 0 & \cdots  \\
0 & M_{11} & M_{12} & 0 & \ddots \\
0 & 0 & M_{22} & M_{23} & \ddots \\
0 & 0 & 0 & M_{33} & \ddots \\
\vdots & \vdots &\vdots &\vdots & \ddots
\end{array}
\right)\ ,
\end{equation}
where each entry is a $(p+1) \times (p+1)$ matrix of the form
\begin{equation}
\label{eq:Mij}
M_{ij}=
\left(
\begin{array}{c c c}
\frac{\partial\beta^{(0)}_{2i}}{\partial {\tilde{\lambda}}^{(0)}_{2j}}&
\cdots &
\frac{\partial\beta^ {(0)}_{2i}}{\partial {\tilde{\lambda}}^{(p)}_{2j}}\\
\vdots &\ddots & \vdots\\
\frac{\partial\beta^{(p)}_{2i}}{\partial {\tilde{\lambda}}^{(0)}_{2j}}&
\cdots &
\frac{\partial\beta^{(p)}_{2i}}{\partial {\tilde{\lambda}}^{(p)}_{2j}}
\end{array}
\right) \, \ ,
\end{equation}
 while $p$ is the highest power of scalar curvature included in the action.
It turns out that,
\begin{equation}
\label{eq:Mzeros}
M_{ij} = 0 \, \, \forall \, i \geq 1, \, \forall \, j<i \, \ ; \qquad
M_{ij} = 0 \, \, \forall \, i, \, \forall \,  j>(i+1) \, \ .
\end{equation}
The various nonzero entries follow the same relations that were observed in
\cite{Gaurav}. In $d$ dimensions they are
\begin{equation}
\label{eq:stab_mat_rel}
M_{ii} = (d-2) i \, {\bf 1}+ M_{00} \, \ ; \qquad
 M_{i,i+1} = (i+1)(2i + 1) M_{01} \ ;
\end{equation}
where
\begin{eqnarray}
\label{eq:M00_M01}
M_{00}&=&
\left(
\begin{array}{c c c c}
\delta M_{\tilde{\lambda}^{(0)}_{0},\tilde{\lambda}^{(0)}_{0} }&
\cdots&\cdots&
\delta M_{\tilde{\lambda}^{(0)}_{0},\tilde{\lambda}^{(p)}_{0} }
\\
\vdots &
\ddots &&
\vdots
\\
\vdots&&\ddots&\vdots
\\
\delta M_{\tilde{\lambda}^{(p)}_{0},\tilde{\lambda}^{(0)}_{0} }&
\cdots&\cdots&
\delta M_{\tilde{\lambda}^{(p)}_{0},\tilde{\lambda}^{(p)}_{0} }
\\
\end{array}\right)
+
\left(
\begin{array}{c c c c c}
-d        &    0      &    0      & \ldots & 0\\
0         &  -(d-2) &    0      &  \ldots & 0\\
 0        &      0    & -(d-4)  &   \ldots & 0 \\
\vdots &  \vdots &     \vdots  &    \ddots &     0\\
   0      &     0      &      0    &    0   & -(d-2p)\\
\end{array}
\right)
\, \ ; \qquad\\
M_{01} &=&
\left(
\begin{array}{c c c c}
\delta M_{\tilde{\lambda}^{(0)}_{0},\tilde{\lambda}^{(0)}_{2} }&
\cdots&\cdots&
\delta M_{\tilde{\lambda}^{(0)}_{0},\tilde{\lambda}^{(p)}_{2} }
\\
\vdots &
\ddots &&
\vdots
\\
\vdots&&\ddots&\vdots
\\
\delta M_{\tilde{\lambda}^{(p)}_{0},\tilde{\lambda}^{(0)}_{2} }&
\cdots&\cdots&
\delta M_{\tilde{\lambda}^{(p)}_{0},\tilde{\lambda}^{(p)}_{2} }
\\
\end{array}\right) \, \ . \qquad
\end{eqnarray}
Using the same arguments as in \cite{Gaurav}, one can prove the above properties
starting from eq. (\ref{eq:Va_beta_s}) neglecting $\partial _t V_a$ and $\partial _t V_a^{\prime}$ on the
right hand side (corresponding to a one-loop approximation). Solving eq. (\ref{eq:Va_beta_s}) beyond
that level would require solving a functional differential equation and would be beyond the scope of this paper.
However, the results presented in the next section suggest that these relations should hold exactly. They
are relations independent of the gauge choice, however the entries
of $M_{00}$ and $M_{01}$ are gauge dependent.

The physical nature of the relations among the eigenvalues can be understood from
the difference between the GMFP and the Gaussian fixed point where also the
gravitational couplings would vanish. At a Gaussian fixed point, the critical
exponents are determined by the mass dimension of the couplings, and therefore
are all spaced by $d-2$.
At the GMFP, the gravitational couplings lead to some corrections to
the critical exponents, but the correction to all exponents is the
same, such that the spacing remains equal to $d-2$.

These relations have important consequences. Because the stability matrix at the
GMFP has the block diagonal structure given by eq. (\ref{eq:stmatform}), its eigenvalues
are just the eigenvalues of the diagonal blocks. Since the diagonal blocks are related
by eqs. (\ref{eq:stab_mat_rel}), the eigenvalues of the various blocks differ only by multiples of $d-2$.
That means if $\rho_0^{(0)},\ldots ,\rho_0^{(p)}$ are the eigenvalues of $M_{00}$, then
all the eigenvalues of $M$ are of the form
\begin{equation}
\label{eq:eng_rel}
\rho_{2i}^{(a)} = \rho_0^{(a)}+(d-2)\, i  \, \ .
\end{equation}
As $M_{00}$ depends only on the couplings
 $\lambda^{(a)}_{0}$, it is enough to include
only these couplings into the action to find all the eigenvalues of the stability matrix.
Therefore, the results for minimally coupled scalar-tensor theory determine the
eigenvalues of the nonminimally coupled scalar-tensor theory. In particular, if one
has calculated the dimension of the UV critical surface of the minimally coupled theory,
one can also predict the dimension of the UV critical surface of the nonminimally coupled theory.

To find all the eigenvectors of the stability matrix it is necessary to know also $M_{01}$.
One can write the eigenvectors as $v=(v_0,v_1,\ldots,v_q)^T$ where each $v_i$ is
itself a $p+1$ dimensional vector. Then the vector $V_0  = (v_0,0,0,\ldots,0)^T$ is an eigenvector
if $v_0$ is an eigenvector of $M_{00}$ which can be seen immediately by multiplying it with $M$.
The eigenvectors of $M$ with the above form are eigenvectors for the eigenvalues
of $M_{00}$ and can therefore be completely determined by just using $M_{00}$.
Thus we note at this point that these eigenvectors
are mixtures of gravitational couplings only, they do not contain any contribution from matter couplings.

Now consider a vector of the form $V_1 =(v'_0,v_1,0,0,\ldots,0)^T$.
Acting on it with $M$, and demanding $V_1$ to be an eigenvector of $M$
corresponding to some eigenvalue $\rho_2^{(a)}$, we obtain two relations,
\begin{equation}
\label{eq:M11_engvec}
M_{00} \, v'_0 + M_{01} \, v_1 = \rho_2^{(a)}\, v'_0 \, \ , \qquad
M_{11} \, v_1 =  \rho_2^{(a)}\, v_1 \, \ .
\end{equation}
The second equation in (\ref{eq:M11_engvec}) tells that $v_1$ is an eigenvector of $M_{11}$.
Now due to equations given in (\ref{eq:stab_mat_rel}) and (\ref{eq:eng_rel}), we note that $v_1 = v_0$.
Determining $v_1$ will then determine also $v'_0$. In the same way one can go on to determine the next eigenvector.
Consider $V_2 =(v''_0,v'_1,v_2,0,\ldots,0)^T$. We then demand it to be a eigenvector of $M$. That means it
should satisfy
\begin{equation}
\label{eq:M22_engvec}
M_{00} \, v''_0 + M_{01} \, v'_1 = \rho_4^{(a)}\, v''_0 \, \ , \qquad
M_{11} \, v'_1 +M_{12} \, v_2 =  \rho_4^{(a)}\, v'_1 \, \ , \qquad
M_{22} \, v_2 =  \rho_4^{(a)}\, v_2 \, \ .
\end{equation}
One notices immediately that $v_2$ is the eigenvector of $M_{22}$, and
using equations in (\ref{eq:stab_mat_rel}) and (\ref{eq:eng_rel}) we conclude that
$v_2 = v_0$. Other equations would determine $v''_0$ and $v'_1$. This process
can be continued to find all the eigenvectors.

We will now illustrate the validity of these results
in various truncations with scalar fields coupled minimally and nonminimally to gravity.

\section{Numerical results}
\label{numericalresults}
\subsection{Nonminimally coupled scalar field}
\label{sectionnonminscalar}

From here on we proceed as in \cite{cpr}. We choose the gauge $\alpha =0$, $\beta\rightarrow \infty$, and $\rho = 0$.
This simplifies the calculation considerably because with that choice several arguments in the FRGE cancel with
each other. The cutoff operators are chosen so that the modified inverse propagator is identical to the
inverse propagator except for the replacement of $z=-\nabla^2$ by $P_k(z)=z+R_k(z)$;
we use exclusively the optimized cutoff functions $R_k(z)=(k^2-z)\theta(k^2-z)$ \cite{Litim}.
Then knowledge of the heat kernel coefficients which contain at most $R^4$ taken from \cite{Avramidi}
is sufficient to calculate all the beta functions.
A further benefit of this choice of cutoff is that the trace arguments will be polynomial in $z$.
This simplifies the integrations in the trace evaluation and is
done in closed form.

Inserting everything into the FRGE and comparing the terms with equal powers of $R$ and
$\phi^2$ on each side of the equation will give a system of algebraic equations for the
beta functions of the couplings $\tilde\lambda_{2i}^{(a)}$. The fixed points of the
flow equations are evaluated and the corresponding critical exponents $\vartheta_{2i}^{(a)}$ are determined.

We carried out the calculation for effective average actions including up to $R^4$ and
up to $\phi^2$ in each potential $V_a$. Such truncations include at most ten couplings.
We find that a GMFP does indeed exist for all these truncations.

\begin{table}
[h]
\begin{center}
\begin{tabular}{|c|r|r|r|r|r|r|r|r|r|r|}\hline
 $p$ & $\tilde \lambda^{(0)}_{0*}$ & $\tilde \lambda^{(1)}_{0*}$ & $\tilde \lambda^{(2)}_{0*}$ & $\tilde \lambda^{(3)}_{0*}$& $\tilde \lambda^{(4)}_{0*}$
\\
\hline
1 & 6.495 & -21.579 & & &\\
2 & 5.224 & -16.197 & 1.834 & &\\
3 & 6.454 & -20.756 & 1.071& -6.474 &\\
4 & 6.354 & -21.342 & 0.792& -6.807 & -3.865\\
\hline
\end{tabular}
\end{center}
\caption{Nonvanishing couplings at the GMFP.
The index $p$ is the highest power of $R$ included in the truncation.
All values are multiplied by a factor 1000.}
\label{table1}
\end{table}
\begin{table}
[h]
\begin{center}
\begin{tabular}{|c|r|r|r|r|r|r|r|r|r|r|}
\hline
 $p$ & $\vartheta'_0$ & $\vartheta''_0$ & $\vartheta^{(2)}_0$ & $\vartheta^{(3)}_0$
& $\vartheta^{(4)}_0$& $\vartheta'_2$ & $\vartheta''_2$ & $\vartheta^{(2)}_2$& $\vartheta^{(3)}_2$& $\vartheta^{(4)}_2$\\
\hline
1 & 2.493 & 2.368 & & & & 0.493& 2.368& & & \\
2 & 1.826 & 2.366 & 21.822 & & & -0.174& 2.366& 19.822 & &\\
3 & 3.077 & 2.524 & 2.033 & -3.852 & & 1.077 & 2.524 & 0.033 & -5.852 &\\
4 & 3.261 & 2.772 & 1.670 & -3.593 & -5.182& 1.261& 2.772& -0.330& -5.593& -7.182\\
\hline
\end{tabular}
\end{center}
\caption{Critical exponents at the GMFP.
The index $p$ is the highest power of $R$ included in the truncation.
Critical exponents are labeled $\vartheta_{2i}^{(a)}$, like the couplings,
but the corresponding eigenvectors involve strong mixing, as discussed in the text.
For each $i$, the first two critical exponents form a complex conjugate pair given by
$\vartheta'_0\pm\vartheta''_0 i$ and $\vartheta'_2\pm\vartheta''_2 i$.
}
\label{table2}
\end{table}
The nonvanishing fixed point values for various truncations are given in table \ref{table1},
the corresponding critical exponents (the negative of the eigenvalues of the stability matrix) in table \ref{table2}.

From the critical exponents one realizes at once several features. Though we carry out the full FRGE calculation
we find that in general the real parts of the critical exponents $\vartheta^{(a)}_{2}$ differ from $\vartheta^{(a)}_{0}$
exactly by two as proven in the one-loop case while the imaginary parts of the critical exponents are unchanged.
This suggests strongly that the relations among the
eigenvalues will also hold at the exact level.
The qualitative and quantitative properties turn out to be very similar to those of the purely gravitational theory.

The inclusion of only four couplings with $a=0,1$ and $i=0,1$
leads to four attractive directions.
The complex critical exponents $\vartheta'_0\pm\vartheta''_0 i$ are expected
from the experience with the Einstein-Hilbert truncation. The existence of a
second pair of complex critical exponents $\vartheta'_2\pm\vartheta''_2 i$
follows from the relation between the eigenvalues given in eq. (\ref{eq:eng_rel}).
These complex conjugate pairs occur also when higher scalar curvature terms are included.

When one includes also $R^2$ couplings, one encounters large positive
critical exponents as known from the calculations in pure gravity
\cite{LauscherR2,cpr,cpr2,Machado:2007ea}. Using eq. (\ref{eq:eng_rel}) one concludes
that one has to go up to power $\phi^{20}$ before encountering a negative
critical exponent, so the critical surface would be twelve dimensional.
But this is a fluke of the $R^2$ truncations due to the anomalously large positive critical exponent.
The situation quickly normalizes when one adds further powers of $R$.

Including $R^3$ couplings, classically one would expect only three positive critical
exponents as the classical mass dimensions of
$\lambda^{(0)}_{0}$, $\lambda^{(1)}_{0}$, $\lambda^{(2)}_{0}$,
$\lambda^{(3)}_{0}$, $\lambda^{(0)}_{2}$, $\lambda^{(1)}_{2}$, $\lambda^{(2)}_{2}$,
and $\lambda^{(3)}_{1}$, are $4$, $2$, $0$, $-2$, $2$, $0$, $-2$, and $-4$ respectively.
Apparently, the FRGE calculation, which includes quantum corrections with
large mixing between the various couplings, produces instead six positive
critical exponents in the $R^3$ truncation. The critical exponent $\vartheta^{(2)}_{2}$
is however very close to zero in consistency with the eigenvalue shift in eq. (\ref{eq:eng_rel}). This tells us
that the truncation with $p=3$ has a six-dimensional UV critical surface for any $i \geq 1$.

With the inclusion of the coupling for the $R^4$ operator whose classical
mass dimension is $-4$, one notices that $0<\vartheta^{(2)}_{0}<2$. Thus
one would expect that including the coupling for the operator $\phi^2 R^4$
with classical mass dimension  $-6$, in consistency with eq. (\ref{eq:eng_rel}),
the critical exponent $\vartheta^{(2)}_{2}$ would be negative, and the critical surface
would be five dimensional. Indeed, the inclusion of those couplings does
make $\vartheta^{(2)}_{2}$ negative, leading to five negative and five positive
critical exponents. One can then say, using eq. (\ref{eq:eng_rel}) in the
truncation $p=4$, that for any $i \geq 1$, the critical surface would be five dimensional.

To illustrate our results we display here the stability matrix for the $R^4$
truncation. The entries in the upper left $5\times 5$ block and in the lower right $5\times 5$ block
are the same except the ones on the diagonals which differ by two. The upper right block
is $M_{01}$, the lower left one contains only zero entries:
\begin{equation}
\label{eq:stmat_num}
\left. M\right|_{\rm GMFP} = \left(
\begin{array}{cccccccccc}
 -0.81 & 1.87 & 0.40 & -1.24 & 0.41 & -0.0057 &
   0.0021 & 0.0011 & -0.00039 & -0.000051 \\
 -8.01 & -6.05 & 2.95 & 2.78 & -1.80 & -0.0031 & -0.0093
   & 0.00083 & 0.0024 & 0.00024 \\
 2.16 & 0.27 & -4.57 & 1.64 & -0.041 & 0.00021 &
   -0.00018 & -0.0032 & -0.00038 &
   -5.55 10^-6 \\
 2.95 & -0.61 & -7.46 & 4.13 & 0.44 & -0.00026 &
   -0.0032 & -0.0098 & -0.0019 & -0.000091 \\
 5.12 & 4.95 & 3.34 & -10.52 & 7.79 & 0.00065 & 0.0021
   & -0.0010 & -0.0071 & -0.00075 \\
 0 & 0 & 0 & 0 & 0 & 1.19 & 1.87 & 0.40 & -1.24 & 0.41
   \\
 0 & 0 & 0 & 0 & 0 & -8.01 & -4.05 & 2.95 & 2.78 & -1.80
   \\
 0 & 0 & 0 & 0 & 0 & 2.16 & 0.27 & -2.57 & 1.64 &
   -0.041 \\
 0 & 0 & 0 & 0 & 0 & 2.95 & -0.61 & -7.46 & 6.13 &
   0.44 \\
 0 & 0 & 0 & 0 & 0 & 5.12 & 4.95 & 3.34 & -10.52 & 9.79
\end{array}
\right)\ .
\end{equation}

The eigenvectors corresponding to the five positive critical exponents in the $R^4$
truncation are given by
\begin{equation}
\label{eq:engvec_R4}
\left(
\begin{array}{c}
-0.2774 \pm 0.2693 i \\
0.8574  \\
-0.1206 \pm 0.0634 i  \\
0.0473 \pm 0.1254 i \\
-0.2202 \pm 0.1746 i  \\
0  \\
 0  \\
 0 \\
 0 \\
 0 \\
\end{array}
\right) \, \ , \qquad
\left(
\begin{array}{c}
(15.381 \pm 5.409 i) \times 10^{-4} \\
(-33.008 \pm 13.931 i ) \times 10^{-4} \\
(4.894 \pm 1.980 i) \times 10^{-4} \\
(-2.535 \pm 1.083 i )\times 10^{-4} \\
(5.437 \pm 8.333 i )\times 10^{-4} \\
-0.2774 \pm 0.2692i \\
 0.8574 \\
 -0.1205 \pm 0.0634 i \\
0.0473 \pm  0.1254 i \\
 -0.2202 \pm 0.1746 i \\
\end{array}
\right) \, \ , \qquad
\left(
\begin{array}{c}
-0.3845 \\
 -0.07586 \\
-0.7103 \\
 -0.5667 \\
-0.1437 \\
0 \\
0 \\
0 \\
0 \\
0\\
\end{array}
\right)\ .
\end{equation}

The first complex conjugate pair of eigenvectors corresponds to the complex conjugate
pair of critical exponents $\vartheta^{\prime}_{0} \pm \vartheta^{\prime \prime}_{0}$
with values $3.2608 \pm 2.7722 i $, while
the second pair of complex conjugate eigenvectors corresponds to the complex conjugate
pair of critical exponents $\vartheta^{\prime}_{2} \pm \vartheta^{\prime \prime}_{2}$ with
values $1.2608 \pm 2.7722 i $. The last eigenvector corresponds to the critical exponent
$\vartheta^{(2)}_{0}=1.6698 $. We note that the eigenvectors corresponding to the eigenvalues
of $M_{00}$, namely the first complex conjugate pair of eigenvectors and
the last one, have the same structure as was described in the previous section,
{\it i.e.} $(v_0,0,0,\ldots,0)^T$, where $v_0$ is determined by just using
$M_{00}$. We note that these eigenvectors do not get mixing from the matter couplings,
but only from the purely gravitational couplings.
Further more, if we look at the eigenvectors corresponding to the
eigenvalues of $M_{11}$, namely the second complex conjugate pair of eigenvectors
in eq. (\ref{eq:engvec_R4}), which has the form $(v'_0,v_1,0,\ldots,0)^T$, we
clearly notice that $v_1 = v_0$, as described in the previous section.
\subsection{Minimally coupled scalar field}
\label{sectionminscalar}
Having verified that the properties of the stability matrix proved at one-loop
level do also hold in the exact calculation, we now
analyze higher order curvature terms retaining
only the couplings $\tilde\lambda^{(a)}_{0}$ corresponding to a
truncation with a minimally coupled scalar field. Then one obtains the
non-Gaussian fixed points and critical exponents
given in tables \ref{table3} and \ref{table4}. We analyze these results and
use them to make predictions for the nonminimal truncation.
\begin{table}
[h]
\begin{center}
\begin{tabular}{|c|l|l|l|r|r|r|r|r|r|r|r|r|}
\hline
 $p$ & $\tilde\Lambda_*$&$\tilde G_*$ & $\Lambda_* G_*$& \multispan9 \hfil $10^3\times$ \hfil
 \vline
\\
\hline
&&& & $\tilde\lambda^{(0)}_{0*}$ & $\tilde\lambda^{(1)}_{0*}$ & $\tilde\lambda^{(2)}_{0*}$ & $\tilde\lambda^{(3)}_{0*}$
& $\tilde\lambda^{(4)}_{0*}$ & $\tilde\lambda^{(5)}_{0*}$ & $\tilde\lambda^{(6)}_{0*}$& $\tilde\lambda^{(7)}_{0*}$& $\tilde\lambda^{(8)}_{0*}$\\
\hline
1& 0.150& 0.923& 0.139& 6.495& -21.579& & & & & & & \\
2& 0.161& 1.228& 0.198& 5.224& -16.197& 1.834& & & & & & \\
3& 0.155& 0.958& 0.149& 6.454& -20.756& 1.071& -6.474& & & & & \\
4& 0.149& 0.932& 0.139& 6.354& -21.342& 0.792& -6.807& -3.865& & & & \\
5& 0.149& 0.932& 0.139& 6.355& -21.339& 0.793& -6.793& -3.854&-0.024& & & \\
6& 0.146& 0.918& 0.134& 6.312& -21.669& 0.586& -7.169& -5.576&-0.537& 2.702 & & \\
7& 0.146& 0.917& 0.133& 6.318& -21.702& 0.534& -6.469& -5.530&-1.979& 2.761 &2.565& \\
8& 0.148& 0.926& 0.137& 6.344& -21.489& 0.678& -5.922& -4.574&-2.074& 1.863 &2.393& 0.829 \\\hline
\end{tabular}
\end{center}
\caption{Position of the FP
for increasing number $p$ of couplings included. The first three columns
give the FP values in the form of cosmological and Newton constant and their
dimensionless product. The values $\tilde\lambda^{(a)}_{0*}$ (and only them) have been rescaled by a factor 1000.}
\label{table3}
\end{table}
\begin{table}
[h]
\begin{center}
\begin{tabular}{|c|r|r|r|r|r|r|r|r|r|}\hline
 $p$ & $\vartheta^{\prime}_{0}$ & $\vartheta^{\prime \prime}_{0}$ & $\vartheta^{(2)}_{0}$ & $\vartheta^{(3)}_{0}$
& $\vartheta^{(4)}_{0}$ & $\vartheta^{(5)}_{0}$ &$\vartheta^{(6)}_{0}$ & $\vartheta^{(7)}_{0}$
& $\vartheta^{(8)}_{0}$\\
\hline
1& 2.493& 2.368& & & & & & & \\
2& 1.847& 2.397& 21.031& & & & & &\\
3& 3.077& 2.524& 2.033& -3.852& & & & &\\
4& 3.261& 2.772& 1.670& -3.593& -5.182& & & & \\
5& 2.777& 2.908& 1.795& -4.176& -4.196 & -6.764 & & &\\
6& 2.841& 2.813& 1.386& -4.000& -3.798 & -5.947 & -8.538 & &\\
7& 2.930& 2.964& 1.312& -4.009& -2.760 & -4.623 & -7.459 & -11.166 &\\
8& 2.331& 2.902& 1.570& -4.063& -0.673 & -7.120 & -7.323 & -9.854 & -11.611\\
\hline
\end{tabular}
\end{center}
\caption{Critical exponents
for increasing number $p$ of couplings included.
The first two critical exponents are a complex conjugate pair of the form
$\vartheta'\pm\vartheta'' i$. The same is the case for the fourth and fifth critical exponent $\vartheta^{(4)}_0\pm\vartheta^{(5)}_0 i$.}
\label{table4}
\end{table}
One observes that the addition of the scalar fields alters the results
for pure gravity in \cite{cpr,cpr2} only by a small amount.
Just as there, the UV critical surface becomes at most three-dimensional,
and fixed point values for the cosmological and the
Newton constant remain very stable. It has to be remarked that for
those two couplings the oscillation in the fixed point value
after the introduction of the $R^2$-term is not as strong as in pure gravity.
Also the critical exponent obtained after the introduction of
the $R^2$-coupling becomes large, but not as large
as in pure gravity. So the addition of the scalar field seems to have
already a little stabilizing effect on the $R^2$-truncation.
The introduction of the $R^4$ and $R^5$-couplings leads to a
second complex conjugate pair of critical exponents as soon as both couplings are included.

Now it is easy to analyze how the dimension of the UV critical surface
changes under the introduction of nonminimal matter couplings.
In general, if a critical exponent $\vartheta^{(a)}_{0}$ is negative then  $\vartheta^{(a)}_{2i}$
will also be negative for all $i>0$. From  table \ref{table4} we see that
$\vartheta^{(a)}_{0} < 0$ for all $a \geq 3 $, thus all $\vartheta^{(a)}_{2i} <0$ for all $a \geq 3$ and $i>0$.
However, since $4>\vartheta^{\prime}_{0} > 2$, using eq. (\ref{eq:eng_rel})
we conclude that $2>\vartheta^{\prime}_{2} > 0$. This means that there are two
more attractive directions. From table \ref{table4} one sees however that
$0<\vartheta^{(2)}_{0}<2$ as soon as $R^4$ is included, thus we do not obtain any other
attractive directions. So compared to \cite{cpr,cpr2} where
a three-dimensional UV critical surface was obtained for pure gravity, interactions
with scalar matter lead to a five-dimensional UV critical surface.

\section{Conclusion}
\label{sectionconclusions}

We have shown that a Gaussian matter fixed point does exist also under the inclusion of higher order
curvature terms and their coupling to scalar fields. We verified that the properties of the stability matrix
proven only at one-loop level hold also in the exact calculations. We exploited these properties to show
the relations between minimally and nonminimally coupled scalar-tensor theory. In particular, we were able
to calculate the critical exponents for the nonminimal scalar tensor theory from those of the minimal
one. The introduction of minimally coupled scalar matter fields gives only slight quantitative corrections
to the fixed point properties of the purely gravitational theory.
The critical exponents again seem to converge with the inclusion of more curvature terms.
The minimally coupled theory produces three positive critical exponents.
We derived that the additional critical exponents in the
nonminimally coupled theory will be the ones of the minimal theory shifted by constant
values. This produces two more positive critical exponents.
From that we can conclude that, in four dimensions, the scalar-tensor theory based
on an action polynomial in scalar curvature and in even powers of scalar field
gives rise to a five-dimensional UV critical surface.

\vspace{0.3cm}

\centerline{\bf Acknowledgements}

\vspace{0.5cm}

We would like to thank R. Percacci for many useful conversations and advice on
the manuscript. C.R. would like to thank Daniel Litim for discussions,
and Gutenberg-University, Mainz, and SISSA, Trieste for hospitality during different stages of this work.

\goodbreak

\end{document}